\documentclass{acm_proc_article-sp}

\begin{document}

\title{Local Popularity based Page Link Analysis}
%
%
%
%
%
\numberofauthors{1} 
%
\author{
%
%
\alignauthor
C Ravindranath Chowdary\\
       \affaddr{AIDB Lab}\\
       \affaddr{IIT Madras}\\
       \affaddr{India}\\
       \email{ravindranathchowdaryc@gmail.com}
}

\maketitle
\begin{abstract}
In this paper we introduce the concept of dynamic link pages. A web site/page contains a number of links to other pages. All the links are not equally important. Few links are more frequently visited and few rarely visited. In this scenario, identifying the frequently used links and placing them in the top left corner of the page will increase the user's satisfaction. This process will reduce the time spent by a visitor on the page, as most of the times, the popular links are presented in the visible part of the screen itself. Also, a site can be indexed based on the popular links in that page. This will increase the efficiency of the retrieval system. We presented a model to display the popular links, and also proposed a method to increase the quality of retrieval system. 
\end{abstract}




\section{Introduction}
Information overload problem is one of the recent hard problems. There is a huge amount of information on the web and lot is being added to it constantly. Organizing the information on a page is a challenging task. While designing a web page (site) for an organization, the designer will organize the information in the form of pages that are linked from main page. But all the information is not equally important, in other words, all the links are not used frequently or each link will be having different frequency. In this paper, we propose dynamic link pages that have a fixed space on the page to accommodate the links that are frequently used in addition to their present location. Note that the frequently used links will also be present at their actual locations. 

\begin{figure}
\label{1}
\centering
\epsfig{file=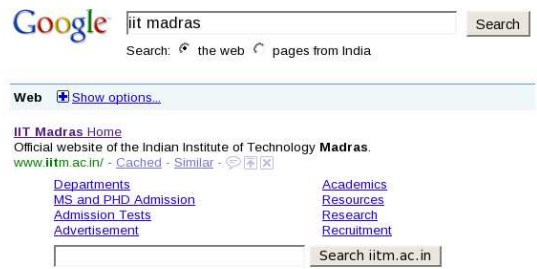, height=2in, width=3in}
\caption{A screen shot of retrieval system for the query ``iit madras''}
\end{figure}

Few of the current day retrieval systems are displaying the frequently visited links of the first retrieved site as shown in Figure \ref{1}. This facility is not available with all the sites and links. In contrast, we propose a model where each site will allocate a fixed space to publish the links that are visited frequently. So, instead of retrieval systems maintaining the log of links, the details can be provided in the site itself. The advantage is two fold. The retrieval systems need not do link analysis within a site. The time spent by a user on visiting a site is minimized as the frequently visited links are readily available in the visible portion of the screen. 

\textbf{Motivation}: With the advancement of Internet, information overload problem has emerged as one of the challenging problems. User is overwhelmed with information. Also, retrieval systems give only ranked list of links. Many a times, user has to go through other links in the page to satisfy his information needs. Also, the links that appear as part of the user query, take the popularity of the site into consideration instead of individual links in that site. This two factors motivated me to work towards a solution. 

\textbf{Contributions}: 
\begin{enumerate}
\item 
Introduced a new methodology of displaying the popular links so as to decrease the time spent by a user after visiting a site.
\item
Given a framework to improve the quality of retrieval systems. 
\item 
Proposed a framework to improve the quality of summaries generated on multiple text documents. 
\end{enumerate}
Based on the usage of links on a page, the importance score can be assigned to the links in the site and improve the performance of retrieval system. This framework is discussed in Section \ref{IR}. Text summarization is one of the solutions that is adopted to overcome the information overload problem. In Section \ref{SUMM}, a framework that improves the performance of a summarization task is outlined. 

\section{Related Work}
World Wide Web has huge amount of information and more is being added to it constantly (Information overload problem). With the advent of Internet, the availability and accessibility of information has increased. As information is made available from multiple sources, it is becoming difficult to a user to go through multiple sites for satisfying his information needs. In this scenario, generating a summary from these multiple sites is of great value.

\textbf{Text Summarization}: Summarization task can be classified based on various criteria. It can be Abstractive - Extractive,  Generic - Query Specific, Indicative - Informative or Single Document - Multiple Documents. Abstractive summaries \cite{Ultra-summarization,SentenceCompression,NaturalLanguageSummaries} deal with generating an abstract by reformulating few sentences whereas extractive summaries \cite{MEAD,ML_ClASSY,QueSTS} involve extracting important sentences from the document(s). 
A summary is generic \cite{MEAD} if it provides overall sense of the document whereas query specific \cite{QueSTS} if it is biased towards a topic. A summary is indicative if it indicates the structure or contents of the document(s) whereas informative if it gives a comprehensive note. Extraction based approaches \cite{Varadarajan, QueSTS, MEAD} give a score to each sentence in the document and the top few sentences are selected as summary. In most of the text summarizers, node scores are calculated by following ideas similar to PageRank \cite{PageRank} and HITS \cite{HITS} and edge scores are calculated based on the amount of similarity between nodes. 

\textbf{Organization of a Web Page}: Wiki page is an example of a well organized structure. In fact, every web page is designed carefully to fulfill the purpose of the site. There are variety of themes that are adopted in designing a web site. The following are the classifications of them: 1) Flat Structure 2) Linked Structure and 3) Mixed Structure. In \textit{Flat Structure}, all the content is made available in the first page of the site itself. User has to scroll up and down in the page to get the information from the site. Links also can be made available on the page to go to any portion of the page. In \textit{Linked Structure}, site will have links from the main page to other pages. Information is segmented and each segment is stored in different page(s). Each segment will have a link from the main page. \textit{Mixed Structure} is the combination of both Flat and Linked structures. Our model works on all these three structures.  

\textbf{Link Analysis by Retrieval Systems}: A page that is part of a popular site will have a different importance when compared to a page that is not part of an important site. One of the main disadvantages of the current scenario is: ``The links that are part of a popular site will be visited more frequently than the links of a site that are not frequently used. If the most relevant information on a topic is present in a link that is not part of an important site and if some information on that topic is present in a link that is part of a popular site then it is highly likely that the link that is part of a popular site will get higher priority.''  

Retrieval systems rank all the pages on the web by considering both the popularity of the site in which they are present and the number of clicks on them. Consider an example: Let there be two sites $A$ and $B$. $A$ is popular than $B$. Two new pages, $a$ and $b$ are added to $A$ and $B$ respectively. Both these new pages have information on the same topic but $b$ is has more valuable information than $a$. In this scenario, $a$ will get a better rank than $b$ (as $A$ is popular than $B$), and only the users who are not satisfied with $a$ will visit $b$. But the number of clicks that $a$ receives will be at least as many as $b$. Therefore $a$ will always be preferred to $b$ by a retrieval system. This is a classic problem with all the retrieval systems. Our model overcomes this problem. Since $A$ is already popular, the number of users visiting it will be more than the visitors who visit $B$. Therefore, $A$ will retain its popularity.  

\section{Link Analysis}
\label{LINK}
We know that all the links in a page will not be visited with the same frequency. In other words, all the links are not equally important. So, we propose a model to analyze the links on a page. For this, each site will have to maintain two types of counters for each link, 1) $History\_Count(i,j)$ and 2) $Recent\_Count(i,j)$. $History\_Count(i,j)$ is the count of number of times link $j$ of site $i$ is visited from the launch of the site $i$. $Recent\_Count(i,j)$ is the count of number of times link $j$ of site $i$ is visited in the recent past \footnote{This value depends on the site. Typically it can be a week.}. In this section, we propose a model to analyse these counters. Analysis is followed by placing few of the popular links in the upper left portion of the home page. 

\begin{figure}
\centering
\epsfig{file=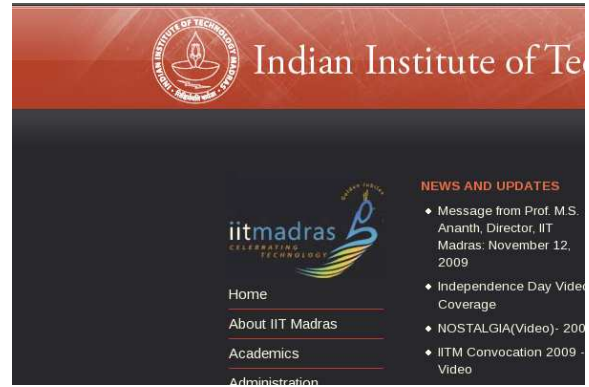, height=2in, width=3in}
\caption{A screen shot of a site before identifying the popular links}
\label{2}
\end{figure}

\begin{figure}
\centering
\epsfig{file=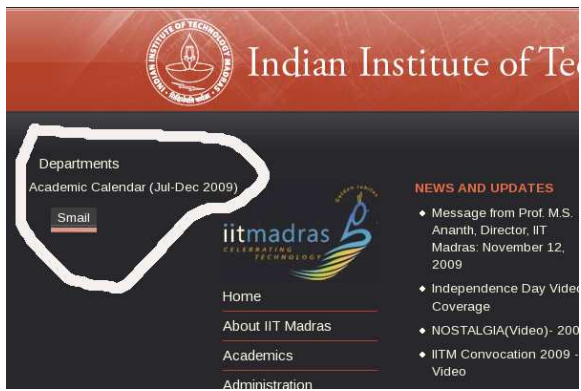, height=2in, width=3in}
\caption{A screen shot of a site after identifying the popular links}
\label{4}
\end{figure}

\textbf{Model}: For each link, a score is computed as the product of $History\_Count(i,j)$ and $Recent\_Count(i,j)$. Top few links are selected based on these scores. Selected links are placed in the upper left corner of the home page. The layout of the site will be as shown in Figure \ref{2}. The layout after identifying popular links is shown in Figure \ref{4}. 

It is assumed that the information of $History\_Count(i,j)$ and $Recent\_Count(i,j)$ is made available by each site as a web service. So, the retrieval system can collect this information from all the sites. 
\section{A model to improve the performance of any IR system}
\label{IR}
Current retrieval systems rank the pages based on their popularity, that is a page will be ranked based on 1) the number of clicks and 2) the popularity of the links that it is connected (both incoming and outgoing). These frameworks fail to capture the importance of a page within its site. The performance of a system can further be improved if the importance of a page within its site is captured. Historical Importance of a link $i$ in site $j$, $HI(i,j)$ is calculated as given in Equation \ref{importanceScore1}. Current Importance of a link $i$ in site $j$, $CI(i,j)$ is calculated as given in Equation \ref{importanceScore2}
\begin{equation}
\centering
\label{importanceScore1}
HI(i,j) = \frac{History\_Count(i,j)}{\sum_{a \in j \bigwedge a \neq i }History\_Count(a,j)} 
\end{equation} 
\begin{equation}
\centering
\label{importanceScore2}
CI(i,j) = \frac{Recent\_Count(i,j)}{\sum_{a \in j \bigwedge a \neq i }Recent\_Count(a,j)} 
\end{equation}  
Using the above two measures, popularity of a link is recalculated. Equations \ref{importanceScore1} and \ref{importanceScore2} are also made part of the popularity calculation. This methodology can be used to improve the quality of the retrieval system. 

\textbf{Model}:  A special weightage is to be given to the links that appear in the top left corner of the site. By doing so, the intentions of users for visiting the site is captured. In other words, the usage of the site (popular links) is obtained. Based on this, retrieval system can recalculate the popularity score of the links and thus improve the performance.  

\section{A framework to improve the summarization task}
\label{SUMM}
Text summarization is one of the useful tasks that grew in popularity recently. Till date, the summary of a site is generated by giving equal priority to all the links in the site. We propose a model where the priority of the links are used while generating the summary. 

\textbf{Framework}: 

Generic Summary Generation: 1) Single site Summarization: Summary of a site is generated by considering only the links that appear on the top left portion of the site. 2) Multiple site summarization: Summary is generated by considering all the links that appear on the top left portion of the sites that are to be summarized. 

Query Specific Summary Generation: 1) Single site Summarization: Summary of a site is generated by considering only the links that appear on the top left portion of the site and have query term(s) in them. 2) Multiple site summarization: Summary is generated by considering all the links that appear on the top left portion of the sites and have query term(s) in them.  

This framework will certainly improve the quality of summarization because, the links that appear on the top left corner are popular and it is highly likely that they will contain information that is useful to users. Also, as the number of links/pages that are to be summarized are getting pruned due to which efficiency will increase. 

\textbf{A framework to improve the site classification}:
A site can be classified based on the popular links in that site i.e., the links in the top left corner. The content of these links can be processed in order to classify the site. Natural Language Processing (NLP) tools are very inefficient and processing few links/pages is efficient when compared to processing all the links in a site. Popular links are in some sense representative pages of their site. Therefore, classification of site is done efficiently.  
\section{Conclusions}
In this paper, a model that also takes into consideration the usage of links within a site is proposed. This can be understood as  giving importance to local (within its site) popularity of a link. In all the frameworks that are proposed in this paper, there is one strong correlation i.e., the models are dynamic. As the priorities of users change within the site, so will the popularity score of the links. In some sense, user's feedback is considered while re-ranking the links. One shortcoming of this paper is that only the frameworks are proposed. 

%


\end{document}